# Precise and reversible band gap tuning in single-layer MoSe$_2$ by uniaxial strain

*Joshua O. Island[1], Agnieszka Kuc[2,3], Erik H. Diependaal[1], Rudolf Bratschitsch[4], Herre S.J. van der Zant[1], Thomas Heine[2,3], Andres Castellanos-Gomez[1,5]*

[1]Kavli Institute of Nanoscience, Delft University of Technology, Lorentzweg 1, 2628 CJ Delft, The Netherlands

[2]Engineering and Science, Jacobs University Bremen, Campus Ring 1, 28779 Bremen, Germany

[3]Wilhelm-Ostwald-Institut für Physikalische und Theoretische Chemie, Universität Leipzig, Linnéstr. 2, 04103 Leipzig, Germany

[4]Institute of Physics, University of Münster and Center for Nanotechnology, D-48149 Münster, Germany

[5]Instituto Madrileño de Estudios Avanzados en Nanociencia (IMDEA Nanociencia), Campus de Cantoblanco, E-28049 Madrid, Spain

**ABSTRACT**. We present photoluminescence (PL) spectroscopy measurements of single-layer MoSe$_2$ as a function of uniform uniaxial strain. A simple clamping and bending method is described that allows for application of uniaxial strain to layered, 2D materials with strains up to 1.1% without slippage. Using this technique, we find that the electronic band gap of single layer MoSe$_2$ can be reversibly tuned by -27 ± 2 meV per percent of strain. This is in agreement with our density-functional theory calculations, which estimate a modulation of -32 meV per percent of strain, taking into account the role of deformation of the underlying substrate upon bending. Finally, due to its narrow PL spectra as compared with that of MoS$_2$, we show that MoSe$_2$ provides a more precise determination of small changes in strain making it the ideal 2D material for strain applications.

## INTRODUCTION

While bulk semiconductors are quite brittle and typically break under strains larger than 1%, two-dimensional (2D) semiconductors can withstand deformations one order of magnitude larger before rupture.[1] This large breaking strength has increased the interest in controlling the electrical and optical properties of atomically thin semiconductors by strain engineering. In the past few years many theoretical works have been devoted to the study of the role of strain on the electronic properties of semiconducting transition metal dichalcogenides.[2-12] Very recently, experimental works employing uniform uniaxial, uniform biaxial, and local uniaxial deformations have determined the role of mechanical strain on the electronic properties of atomically thin $MoS_2$ and $WSe_2$ but the strain tunability of other members of the dichalcogenides family remains unexplored.[13-17]

Here we employ photoluminescence (PL) spectroscopy to probe the changes in the electronic band structure of atomically thin $MoSe_2$ by uniaxial strain. Through a simple clamping and bending technique, we measure reproducible strains in $MoSe_2$ up to 1.1%. Additionally, we directly compare the photoluminescence spectra of $MoSe_2$ to that of $MoS_2$ to show that the narrow photoluminescence peak (smaller full width half maximum) allows for higher accuracy in determining small changes in strain. Upon uniform application of strain, we find that the energy of the direct band gap transition reduces linearly with strain by -27 ± 2 meV per percent of strain for single-layer $MoSe_2$. We carry out density-functional calculations to further support our experimental findings. By taking into account the Poisson's ratio of the underlying substrate, we calculate a band gap shift of -32 meV per percent of strain.

**SAMPLE FABRICATION AND BENDING APPARATUS**

Slippage of the flakes during the straining/relaxing cycles is a well-known problem in uniaxial straining experiments on 2D materials that severely affects the reproducibility of the results. In conventional substrate bending experiments on graphene or $MoS_2$, strain levels of ~1% can be reliably achieved without suffering from slippage. Clamping the flake to the substrate with deposited metal electrodes has proven to be an effective method to solve the slippage issue to a great extent. In fact, Conley et al. have shown reliable uniaxial straining of $MoS_2$ up to 2.2% by evaporating metallic bars onto the flake.[15] However, this method typically requires extra steps of cleanroom fabrication. If a shadow mask is employed instead of lithography, a careful alignment of the mask is still needed.

In this work we fabricate single-layer MoSe$_2$ samples by mechanical exfoliation of bulk crystals onto a thin polydimethylsiloxane (PDMS) substrate (Fig. 1(a)). Subsequently, the face of the PDMS substrate containing thin flakes is gently placed, face-down, onto a flexible polycarbonate substrate, sandwiching the flakes between the two layers. Special care is taken to place the small PDMS film on the central part of the polycarbonate strip to prevent built-in strain before measurements and for an accurate determination of the applied strain. The PDMS film acts as a clamp to secure the flake during straining. Figure 1(b) shows a transmission mode optical image of a MoSe$_2$ flake sandwiched between the PDMS and polycarbonate substrates (see inset of Fig. 1(b)). Single-layers can be easily distinguished from multilayer counterparts because of their strong photoluminescence yield due to the direct band gap nature of monolayer MoSe$_2$ in contrast to multilayered flakes that are indirect gap semiconductors. See the Supplemental Information (SI) for a comparison between the photoluminescence spectra of the single layer portion of this flake (top-most part) and the bilayer region (directly below the single layer). The polycarbonate substrate is then loaded into a custom made, two-point bending apparatus shown in Fig. 1(c) and secured between two screw-posts. The moveable plateaus of the apparatus (arrow at the right-side of in Fig. 1(c)) allow full control over the bend of the polycarbonate substrate. Given the thickness (*t*) of the substrate (0.8 mm) and an estimation of the radius of curvature (*R*) for a particular strain (see SI for details), the strain can be estimated by $\epsilon = t/2R$.[15, 18]

**RESULTS**

Figure 2(a) shows PL spectra measured on monolayer MoSe$_2$ and MoS$_2$ flakes for direct comparison at strain levels of 0% and 1%. The spectra acquired for the relaxed MoSe$_2$ and MoS$_2$ samples (red curves in Figure 2(a) and inset of 2(a), respectively) agree with those reported in the literature.[19-22] Specifically, the prominent peak at 662 nm (782 nm), determined from a Lorentzian fit, corresponds to a direct transition at the K point, giving an optical band gap of 1.59 eV (1.87 eV) for MoSe$_2$ (MoS$_2$).[19-22] It can be seen that PL peak for MoS$_2$ (FWHM of 46 nm) is much broader than that of the MoSe$_2$ peak (FWHM of 22 nm). Upon increasing higher uniaxial strain of 1% the PL peaks shift towards lower energy (red shift). As the exciton binding energy in transition metal dichalcogenides is expected to be nearly independent of the uniaxial strain[23] this shift of the PL emission can be directly correlated with a reduction of the band gap in the monolayer flakes. While the shift in the PL peak for MoSe$_2$ is quite clear, shifting more than one FWHM, that for MoS$_2$ is relatively smaller compared with the width of the peak. This suggests MoSe$_2$ as a superior material in strain applications where precise measurements of small variations in strain are required.

In order to verify that slippage is not affecting the measurements, we subjected a characteristic single layer MoSe$_2$ device to several straining/relaxing cycles. Figure 2(b) shows the peak center, from a fit, for the MoSe$_2$ flake for repeated cycles of straining and relaxing. The PL emission reproducibly shifts from ~782 nm a uniaxial tensile strain of 0% to higher wavelengths for strains of 0.7%, 0.9%, and 1.1%. Between each cycle, the PL emission peak always comes back to the same value indicating that the flake does not slip during the measurement. By repeating this measurement at increasingly high strain levels, we determine the threshold strain value before slippage starts to play an important role. We have found that for strains higher than 1.1 % these measurements are not reproducible anymore and thus we are limited to a range of strains below this threshold.

We now turn to the change in the band gap of single-layer MoSe$_2$ for given strains up to 1.1% using the described bending apparatus. Figure 3(a) shows the shifts in the PL emission peak for a single-layer MoSe$_2$ flake for progressively increasing strain levels. The PL emission steadily shifts toward lower energies, indicating a reduction of the band gap for higher strains. Figure 3(b) shows the change in the band gap energy for two devices. Device 1 corresponds to the PL spectra in Figure 3(a). The change in the band gap per % of strain is extracted from the slope of a linear fit to the data for both devices. We measure a change of -27 ± 2 meV in the bad gap energy per percent strain. While reported values for the band gap change in $MoS_2$ are higher (~45 meV)[15], as pointed out earlier, the peak widths are broader making small variations in strain difficult to resolve.

We have employed Density-Functional Theory (DFT) to calculate the band structure of monolayer $MoSe_2$ at different strain levels (see Materials and Methods for details). Figure 3(c) shows the band structure for single-layer $MoSe_2$. We calculate a band gap of 1.35 eV including spin-orbit coupling. This value, lower than our experimentally measured value of 1.58 eV, is expected as the PBE functional is well known to underestimate the band gap energy. However, our comparative conclusions of the strained systems hold, as the band gap underestimation due to the PBE functional is the same in all cases studied here. Figure 3(d) shows the band structure at a strain of 1.5% for armchair and zigzag strain directions which show similar changes in the band gap (see SI for band gap values at strains from 0% to ~2% for uniaxial strain in both directions and biaxial strain). We calculate a linear change in the band gap of -47 and -48 meV per percent of strain for the armchair and zigzag directions, respectively. Considering this deviation from the experimental shift of -27 meV per percent of strain, it is important to note the strain properties of the polycarbonate substrate itself. The polycarbonate substrate has a Poisson's ratio of ~0.37[24] at room temperature which means that an application of 1 % strain along the long side of the substrate (see Figure 1(c)) results in a contraction of 0.37 % along the short side. Taking this effect into account and applying a perpendicular contraction of 0.37% for 1% uniaxial strain results in a linear band gap shift of -32 meV per percent strain (see SI for linear trend), improving substantially the agreement with the experimental result above.

**CONCLUSION**

In summary, we have observed a red-shift of the PL emission of single-layer $MoSe_2$ subjected to uniform uniaxial tensile strain, corresponding to a strain modulation of the bad gap. A simple technique is described to clamp the single-layer flakes to a bendable polycarbonate substrate and apply reproducible strains up to 1.1% without flake slippage. We find that the PL peak of $MoSe_2$ is much sharper than MoS$_2$ suggesting that the material would be better suited for applications of precise band-gap tuning. The experimental strain tunability of monolayer MoSe$_2$ is found to be -27 ± 2 meV per percent of strain. The measured shift of the PL upon uniaxial strain is in good agreement with DFT calculations that predict a reduction of the band gap of -32 meV per percent of strain taking into account the Poisson's ratio of the underlying substrate. The possibility to tune the PL emission in combination with the bright and narrow PL peak of single-layer MoSe$_2$ opens opportunities to use this material for tunable optoelectronic applications.

## MATERIALS AND METHODS

**Synthesis and characterization:** Bulk MoSe$_2$ material was grown by the vapor phase transport method.[25] X-ray diffraction was performed to confirm the 2H- polytype of the MoSe$_2$ single cyrstals.[21] Raman spectroscopy (not shown) and photoluminescence measurements were performed (Renishaw in via) in a backscattering configuration excited with a visible laser light ($\lambda$ = 514 nm). Spectra were collected through a 100× objective and recorded with a 1800 lines mm$^{-1}$ grating providing a spectral resolution of ~1 cm$^{-1}$. To avoid laser-induced heating and ablation of the samples, all spectra were recorded at low power levels P~500 μW and short integration times (~1 s). Photoluminescence measurements however require longer integration times (~60–180 s).

**Calculations:** All calculations were carried out using density-functional theory (DFT) with the PBE[26] exchange-correlation functional, with London dispersion corrections as proposed by Grimme[27] and with Becke and Johnson damping (PBE-D3(BJ)) as implemented in the ADF/BAND package.[28, 29] Local basis functions (numerical and Slater-type basis functions of valence triple zeta quality with one polarization function (TZP)) were adopted, and the frozen core approach (small core) was chosen. All calculations included the scalar relativistic corrections within the Zero Order Regular Approximation (ZORA).[30-33] We have fully optimized the MoSe$_2$ monolayer (atomic positions and lattice vectors). The optimized lattice parameter of $a$ = 3.322 Å was obtained for the hexagonal representation, in a good agreement with experimental data ($a$ = 3.288 Å)[34]. The atomic positions of MoSe$_2$ monolayer were further reoptimized for a given uniaxial or biaxial tensile strain. Electronic band gaps were obtained both from the ZORA calculations as well as from the simulations with the spin-orbit coupling (SOC). The $k$-point mesh over the Brillouin zone was sampled according to the Wiesenekker-Baerends scheme,[35] where the integration parameter was set to 5, resulting in 15 $k$-points in the irreducible wedge. The calculated band gap of MoSe$_2$ monolayer is 1.46 and 1.35 eV from the ZORA and SOC calculations, respectively.

**FIGURES**

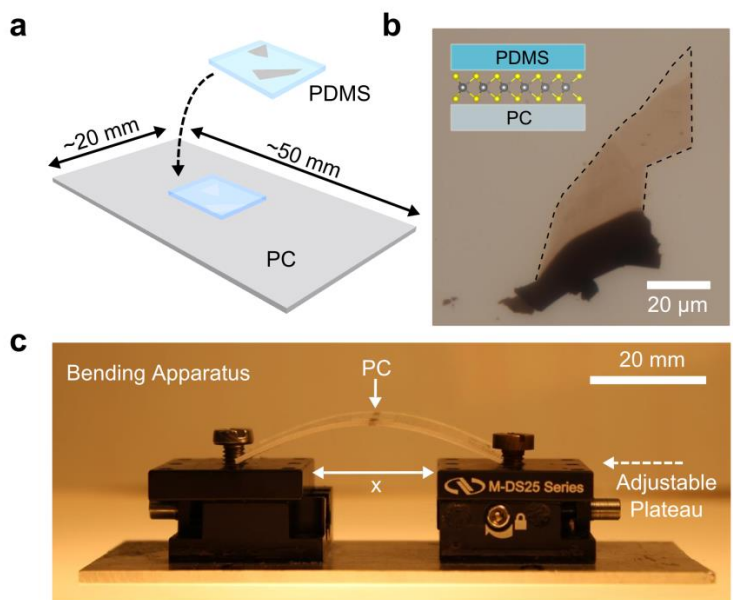

FIG. 1. (a) A viscoelastic (PDMS) stamp with exfoliated MoSe$_2$ is placed upside down onto a polycarbonate (PC) substrate to create a PC-MoSe$_2$-PDMS stack. (b) Optical transmission image of a MoSe$_2$ flake sandwiched between PC and PDMS. The inset shows a cartoon of the sandwich for a single-layer flake. (c) Optical image of the apparatus used to bend the sandwich and apply strain to the MoSe$_2$ flake.

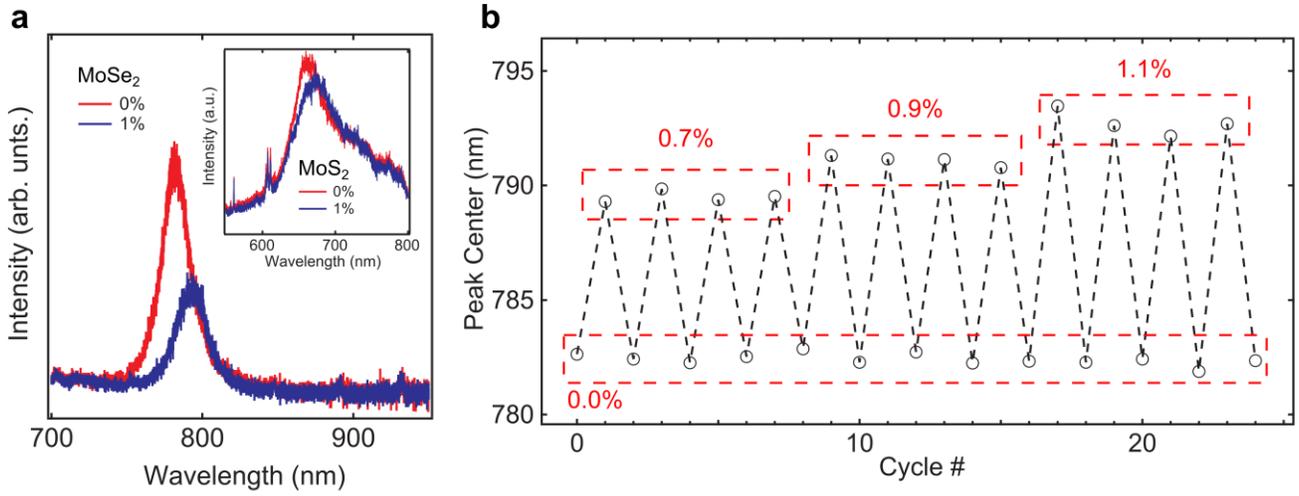

FIG. 2. (a) PL spectra for a single layer MoSe$_2$ flake at 0% and 1% strain. The inset shows the PL spectra for a single-layer MoS$_2$ flake at 0% and 1% strain. Note that the wavelength scale is the same width for both plots showing clearly the difference in the FWHM between the MoSe$_2$ PL peak and the MoS$_2$ PL peak. (b) Center of the PL peak for single-layer MoSe$_2$ as a function of strain for several straining cycles. PL shifts for strains up to 1.1% are reproducible using the simple clamp and bend apparatus in Figure 1(c).

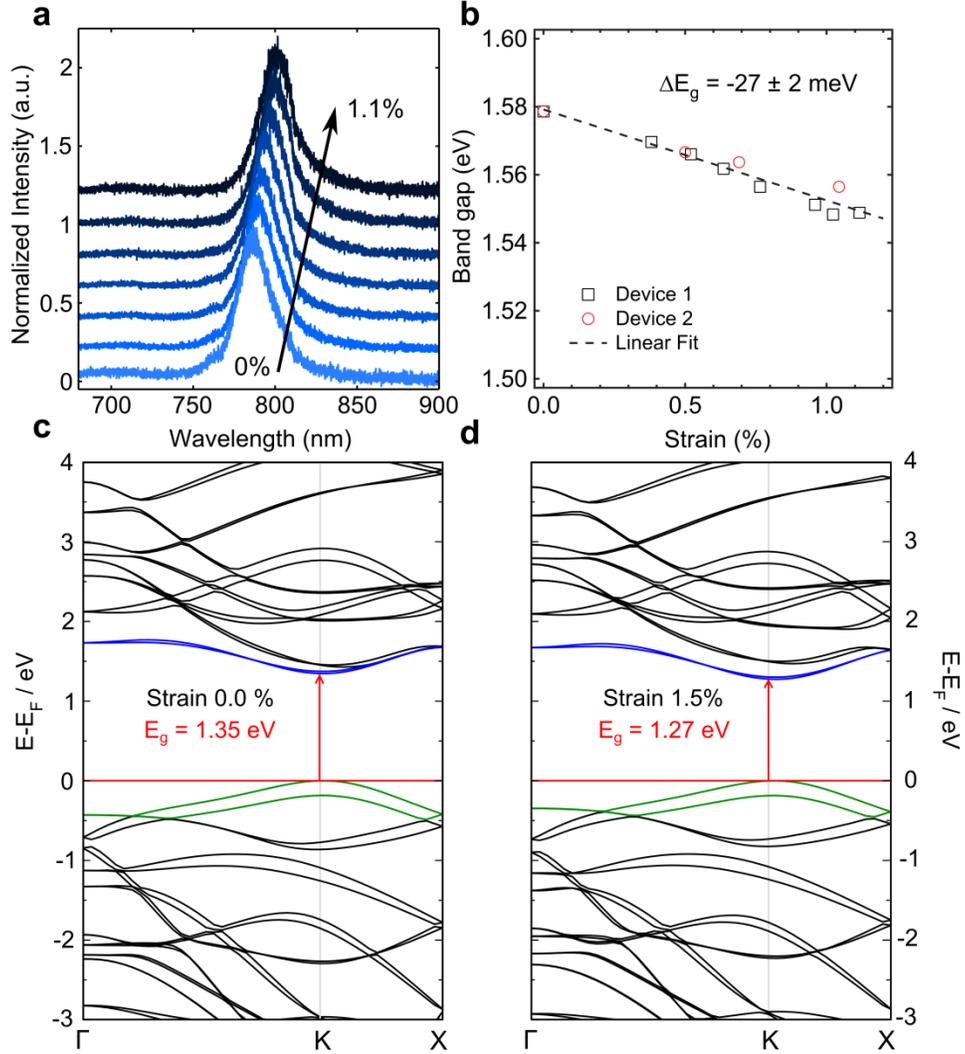

FIG. 3. (a) PL spectra for a single-layer MoSe$_2$ flake for increasing strains up to 1.1%. (b) Change in band gap for two single-layer samples (device 1 is from the spectra in (a)). (c) Calculated band structure of single-layer MoSe$_2$ including spin-orbit coupling at 0% strain and (d) at 1.5% uniaxial strain.

**ACKNOWLEDGMENTS**

The authors would like to thank Gary A. Steele for helpful discussions. This work was supported by the Dutch organization for Fundamental Research on Matter (FOM) and by the Ministry of Education, Culture, and Science (OCW). A.C.-G. acknowledges financial support by the European Union through the FP7-Marie Curie Project PIEF-GA-2011-300802 ("STRENGTHNANO") and by the Fundacion BBVA through the fellowship "I Convocatoria de Ayudas Fundacion BBVA a Investigadores, Innovadores y Creadores Culturales" (Semiconductores Ultradelgados: hacia la optoelectronica flexible) and from the MINECO (Ramon y Cajal 2014 program, RYC-´ 2014-01406) and from the MICINN (MAT2014-58399-JIN). The Deutsche Forschungsgemeinschaft (HE 3543/18-1) and the European Commission (FP7-PEOPLE-2012-ITN MoWSeS, GA 317451) are acknowledged.

# Supplemental Information

# Precise and reversible band gap tuning in single-layer MoSe$_2$ by uniaxial strain

*Joshua O. Island, Agnieszka Kuc, Erik H. Diependaal, Rudolf Bratschitsch, Herre S.J. van der Zant, Thomas Heine, Andres Castellanos-Gomez*

**Supporting Information Contents**
1. Mono- bilayer spectra
2. Uniform strain estimation
3. Band gap change with strain (DFT calculations)

1. **Mono- bilayer spectra**

Figure S1(a) shows an optical transmission image of the same flake as described in the main text. The monolayer and bilayer locations of the flake are indicated. The spectra for each region can be seen in Figure S1(b). The monolayer PL intensity is much larger than the bilayer region, as expected.

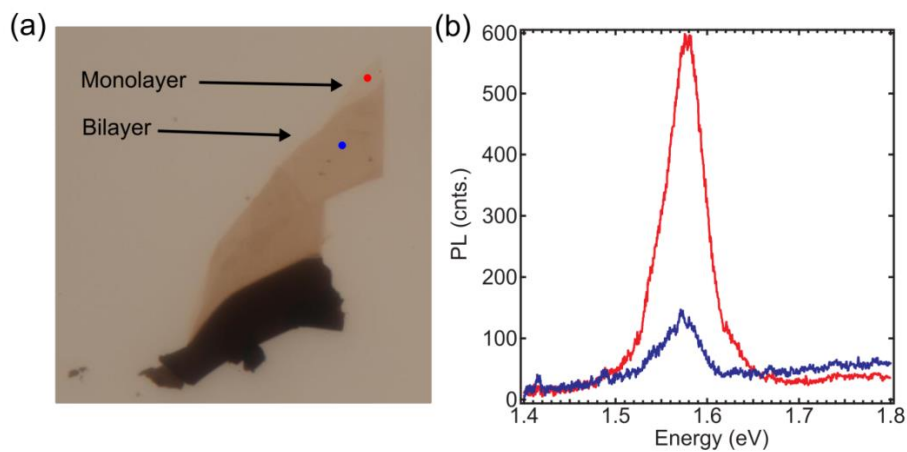

**Figure S1**: (a) Transmission mode optical image of a MoSe$_2$ flake with single, bilayer, and multilayer regions. (b) PL spectra for the single layer portion in (a) (red dot) and bilayer portion (blue dot).

2. **Uniform strain estimation**

Strains ($\epsilon$) are calculated given the thickness (*t*) of the substrate and the radius of curvature (*R*) of the substrate while strained, $\epsilon = t/2R$. The radius of curvature can be simply estimated given an optical image of the strained substrate, such as those in Figure S2(b-d). For more efficient calculation of the strain, given a simple measureable quantity (the distance between the movable plateaus) we wrote a script that uses the bisection method to calculate the radius of curvature given the chord length (from tip to tip) of the substrate arc. The chord length is measured using calipers to measure the distance between the plateaus (See Figure 1(c) of the main text). Figure S2(e) shows an output plot of this script for a sample substrate with a length of 46.3 mm.

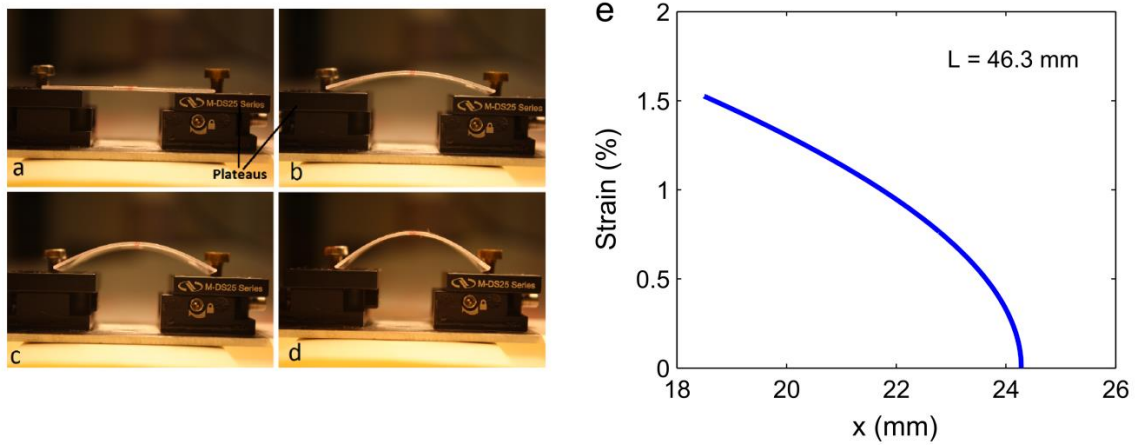

**Figure S2**: (a-d) Optical images of the bending apparatus for varying strains. Strain estimation for a given distance, x, between the platforms of the bending apparatus.

3. **Band gap change with strain (DFT calculations)**

Figure S3 shows the calculated band gap (ZORA and with spin-orbit coupling, ZORA + SOC, see Materials and Methods of the main text). For the armchair and zigzag directions the band gap changes by -47 and -48 meV per percent of strain, respectively (black and red curves in Figure S3). For biaxial strain the band gap changes by -87 meV per percent of strain (green curves in Figure S3). When we take into account the Poisson's ratio of the substrate of 0.37 % for 1% strain, the band gap trend is much shallower, showing a band gap change of -32 meV per percent of strain, in agreement with our experimental strain results.

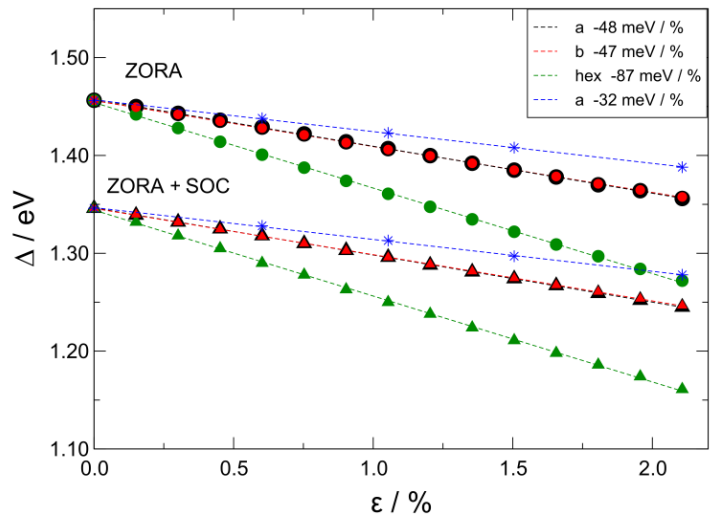

**Figure S3**: Calculated change in band gap for strain up to ~2%. The red and black curves show the trends for the armchair and zigzag directions. The green curve shows the trends for the biaxial strain. The blue curve shows the trend while taking into account the effect of the substrate (see main text).

14